\documentclass{ws-procs9x6-cpt19}

\usepackage{amsfonts}
\usepackage{amsmath}
\usepackage{graphicx}
\usepackage{color}
\usepackage{amssymb}

\begin{document}

\newcommand{\refeq}[1]{(\ref{#1})}
\def\etal {{\it et al.}}

\title{Explicit Spacetime-Symmetry Breaking in Matter:\\
the Reversed Vavilov--\v{C}erenkov Radiation
}

\author{O.J.\ Franca, L.F.\ Urrutia, and Omar Rodr\'\i guez-Tzompatzi}

\address{High Energy Physics Department, Universidad Nacional Aut\'onoma de M\'exico,\\
CDMX, 04510, Mexico}

\begin{abstract}
We show 
that reversed Vavilov--\v{C}erenkov radiation 
occurs simultaneously with the forward output 
in naturally existing materials 
when an electric charge moves with a constant velocity 
perpendicular to the planar interface 
between two magnetoelectric media. 
Using the Green's function in the far-field approximation 
we calculate the angular distribution of the radiated energy per unit frequency 
obtaining a non-zero contribution in the backward direction.
\end{abstract}

\bodymatter

\section{Introduction}

As a motivation for this work 
let us recall that spacetime-dependent couplings have been identified 
as possible sources of fundamental Lorentz-invariance violation.\cite{KLP}  
Also, 
let us consider the photon sector of the Standard-Model Extension\cite{KB}
\begin{equation} 
{\cal L}=-\frac{1}{4} F_{\mu\nu}F^{\mu\nu}- \frac{1}{4}(k_{F})_{\kappa \lambda\mu\nu }F^{\kappa\lambda}F^{\mu\nu}+\frac{1}{2}(k_{AF})^\kappa \epsilon_{\kappa \lambda\mu\nu} A^\lambda F^{\mu\nu}.
\label{LSMEED}
\end{equation}
A term that is missing in the Lagrange density \refeq{LSMEED} 
is $(k_{F})_{\kappa \lambda\mu\nu }=\theta \epsilon_{\kappa \lambda\mu\nu}$, 
with the axion coupling $\theta$ being a constant, 
simply because this contribution is a total derivative. 
Nevertheless, 
if we promote the constant $\theta$ to $\theta(x)$ 
we are able to describe the effective electromagnetic response 
of a host of interesting physical phenomena 
where spacetime symmetries are expected to be explicitly broken. 
In this work 
we deal with axion electrodynamics 
\begin{equation}
{\cal L}=-\frac{1}{4} F_{\mu\nu}F^{\mu\nu}- \frac{\alpha}{32 \pi^2} \, \vartheta(\mathbf{x}) \epsilon_{\kappa \lambda\mu\nu }F^{\kappa\lambda}F^{\mu\nu}, 
\label{LAG}
\end{equation}
which describes the electromagnetic response of magnetoelectric materials 
like topological insulators\cite{TI} and  Weyl semimetals,\cite{WS} 
for example.  
Here we summarize the discovery of reversed Vavilov--\v{C}erenkov radiation (RVCR)
in naturally existing magnetoelectrics, 
reported in Ref.\ \refcite{URRU}.

In 1968 Veselago \cite{Veselago} proposed 
that RVCR 
could be produced in  materials characterized 
by negative permittivity $\epsilon$ and permeability $\mu$.  
In this case 
a radiation cone is observed in the backward direction 
with respect to the velocity of the  propagating charge. 
Since such materials  are not found in nature, 
this proposal gave a major thrust to the design and construction of metamaterials  
that would provide the required properties 
in a given range of frequencies.\cite{Pendry}

\section{RVCR in axion electrodynamics} 

We consider two semi-infinite media 
separated by the interface $\Sigma$ at $z=0$ 
with the axion field  
$\vartheta (z)=H (z)\vartheta _{2}+ H(-z)\vartheta _{1}$ 
and  $\vartheta _{1}$, $\vartheta _{2}$ constants, 
where $H(z)$ is the Heaviside function.   
In order to  suppress the transition radiation 
we take $\epsilon _{1}=\epsilon _{2}=\epsilon $ 
as a first approximation. 
We also set $\mu_1=\mu_2=1$.
The modified Maxwell equations 
resulting from the Lagrange density \refeq{LAG} 
are those corresponding to a normal material medium 
with constitutive relations 
$\mathbf{D}=\epsilon \mathbf{E}-\alpha \vartheta(z) \mathbf{B}/\pi $  
and $\mathbf{H}=\mathbf{B}+\alpha \vartheta(z) \mathbf{E}/\pi $.
These equations introduce field-dependent effective charge and current densities 
with support only at the interface $\Sigma$, 
which produce the magnetoelectric effect,\cite{TI} 
being governed by the parameter 
${\tilde \theta}= \alpha (\vartheta_2-\vartheta_1)/\pi$.
To calculate the  radiation fields 
we rewrite the modified Maxwell equations 
in terms of the potential $A_\mu$ $(c=1)$ 
and introduce the Green's function (GF) 
$G_{\;\;\sigma }^{\nu }(x,x^{\prime })$ 
by setting $A^{\mu }(x)=\int d^{4}x^{\prime }G_{\;\,\nu }^{\mu }({x},{ x}^{\prime} )J^{\nu}( { x}^{\prime })$, 
obtaining 
\begin{equation}
\hspace{-.1cm}\left(\left[\Box^{2}\right]_{\;\;\nu }^{\mu }-\tilde{\theta}\delta
(z)\varepsilon_{\quad \;\nu }^{3\mu \alpha }\partial _{\alpha }\right)
G_{\;\;\sigma }^{\nu }(x,x^{\prime })= 
4\pi \eta _{\;\;\sigma }^{\mu }\delta^4
(x-x^{\prime }).\label{Eq GF temporal}
\end{equation}
Let us  remark 
that the resulting $\theta$-dependent contribution to the GF 
is a function of  $(|z|+|z'|)$, 
instead of $ |z-z'|$ as it is in the normal case. 
This is the mathematical origin of the RVCR.
The required GF in the far-field approximation 
includes integrals of rapidly oscillating functions 
and its leading contribution is obtained 
by the stationary-phase approximation. 

Now, 
let us  consider a charge $q$ 
moving at a constant velocity $v\mathbf{\hat{u}}$ 
with charge and current densities 
$\varrho (\mathbf{x}^{\prime };\omega )=\frac{q}{v}\delta (x^{\prime })\delta(y^{\prime })e^{i\omega \frac{z^{\prime }}{v}},\quad   \mathbf{j}(\mathbf{x}
^{\prime };\omega) =   \varrho  v\mathbf{\hat{u}}$. 
Here, 
${\mathbf {\hat u}}$ is the unit vector 
perpendicular to the interface, 
directed from region 1 to region 2.
We henceforth assume $v>0$ 
and consider the motion in the interval 
$z\in(-\zeta ,\zeta )$, 
with $\zeta \gg v/\omega$. 
In the far-field approximation 
we calculate the electric field $\mathbf{E}$ 
starting from the potential $A^{\mu }$ 
and using the corresponding approximation of the GF 
obtained from  Eq.\ \refeq{Eq GF temporal}. 
Then, 
we are able to calculate the spectral distribution (SD) of the radiation 
$d^{2}E/d\omega d\Omega=(\mathbf{E}^{\ast }\cdot \mathbf{E}) \,  nr^{2}/4\pi ^{2}$ 
in the limit $r \rightarrow \infty$. 
The main point here is 
that the following integrals, 
which result from the convolution of the GF with the sources,
\begin{eqnarray}
\mathcal{I}_{1}(\omega,\theta )&=&\int_{-\zeta }^{\zeta }dz^{\prime }e^{i%
\frac{\omega z^{\prime }}{v}(1-vn\cos \theta )}=\frac{2\sin\left(\zeta\Xi_{-}\right) }{\Xi_{-} },  \label{I1 function} \\
\mathcal{I}_{2}(\omega,\theta )&=&\int_{-\zeta }^{\zeta }dz^{\prime
}e^{i n \omega |z^{\prime }\cos \theta |+i\omega\frac{z^{\prime }}{v}}
= \frac{\sin ( \zeta
{\tilde \Xi}_{-}) }{{\tilde \Xi}_{-}}+
2i \frac{\sin ^{2}( \frac{\zeta }{2}{\tilde \Xi}_{-} ) }{ {\tilde \Xi}_{-} },
\label{I22 function}
\end{eqnarray}%
enter in the expression for the electric field. 
The notation is $\Xi_{-}= \frac{%
\omega }{v}\left( 1 - vn\cos \theta \right)$ and  
$\tilde{\Xi}_{-}= \frac{%
\omega }{v}\left( 1 - vn|\cos \theta | \right)$. 
More importantly, 
in the limit $\zeta \gg v/\omega$,  
($\zeta \rightarrow \infty$), 
the right-hand side of Eqs.\ \refeq{I1 function} and \refeq{I22 function} 
yields expressions like $\sin(\zeta\,\rho N)/(\rho N)$ 
which behave as $\pi \delta(\rho N)$. 
This delta-like behavior means 
that the contributions of the electric field 
to the SD are nonzero only when 
(i) $ 1\pm vn \cos \theta =0$ and/or 
(ii) $ 1\pm vn|\cos \theta |=0$. 
Following our conventions, 
the case (i) yields the standard \v{C}erenkov condition $\cos \theta_C= 1/( n v)$. 
The second case (ii) opens up the possibility 
that $\cos \theta$ is negative 
yielding the reversed \v{C}erenkov cone 
at $\theta_{\rm R}=\pi-\theta_C$. 
In other words, 
the terms containing $\tilde{\Xi}_{-}$ 
make possible the production of radiation 
in the backward direction. 
The SD of such radiation is suppressed with respect to the forward output, 
but nevertheless it is different from zero.

To illustrate our results 
we choose  medium 2 as the topological insulator ${\rm TlBiSe}_2$, 
with $n_2=2$ prepared in such a way that ${\tilde \theta}= 11 \alpha $, 
together with a normal insulator in medium 1 having  $n_1=n_2$. 
We consider the radiation 
emitted at the average frequency of $\omega=2.5\,$eV ($500\,$nm) 
in the \v{C}erenkov spectrum.  
The SD of the total radiation is shown in Fig.~\ref{FULLCHE}. 
In the left panel of the figure 
we plot a zoom in the backward direction 
showing the onset of the RVCR 
and making evident the high suppression of this new contribution 
with respect to the forward Vavilov--\v{C}erenkov radiation.
The details of the above calculations 
are reported in Ref.\ \refcite{URRU}.

\section{Summary}

\begin{figure}
\begin{center}
\includegraphics[scale=0.33]{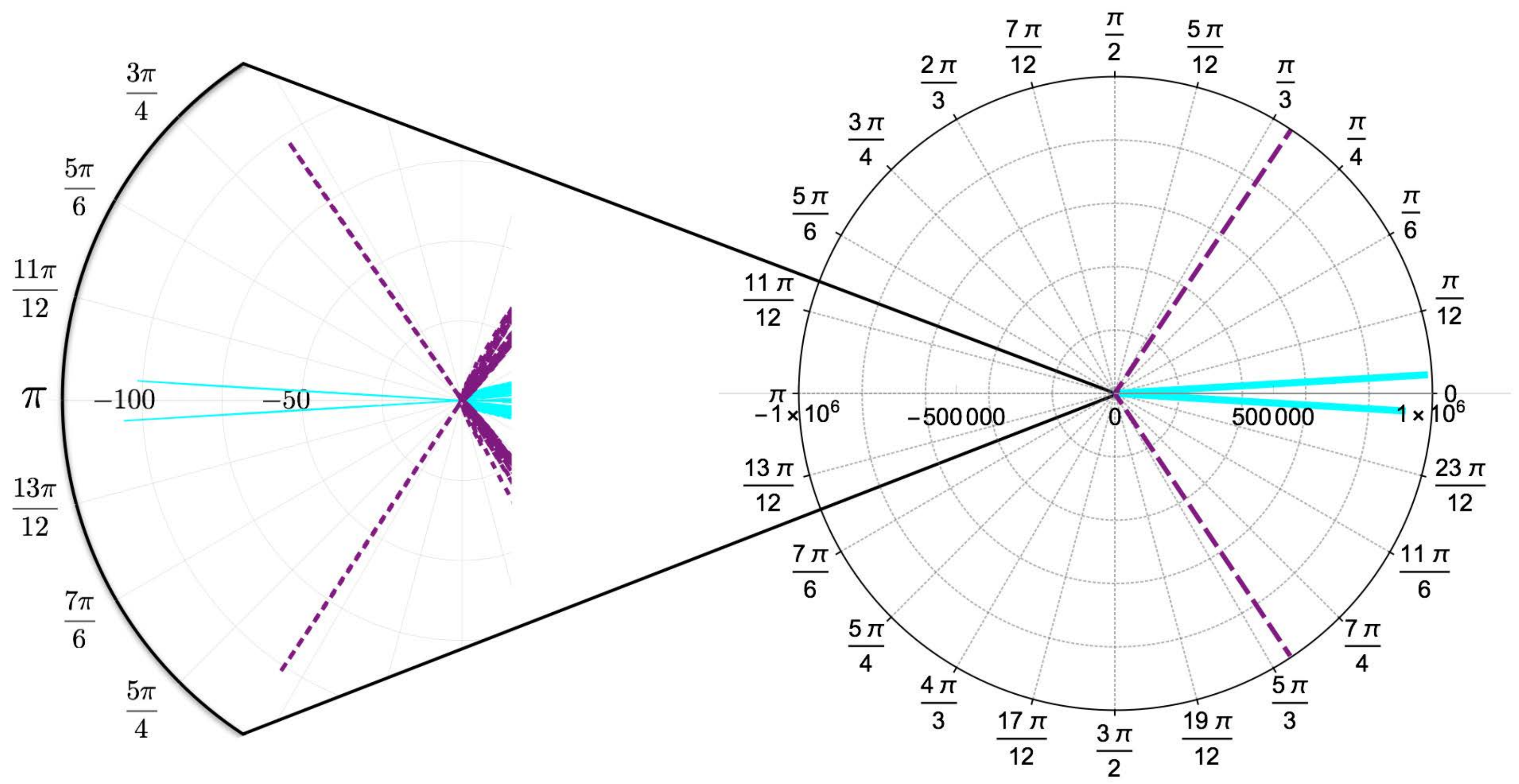}
\end{center}
\caption{Polar plot of the SD for the full Vavilov--\v{C}erenkov radiation 
when $n_1=n_2=2$,  
$\omega=2.5\,$eV, 
and ${\tilde \theta}=11 \alpha$, 
taken from Ref.~\protect\refcite{URRU}. 
The dashed line corresponds to $v=0.9$ and $\zeta=343$ eV$^{-1}$, 
and the solid line to $v=0.501$ and $\zeta=4830\,$eV$^{-1}$. 
The scale of the polar plot is  arbitrary 
and runs from $0$ to $10^6$. 
The left panel in the figure is a zoom of the SD 
in the backward direction 
showing the onset of the RVCR. 
Here, 
the radial scale goes from zero to $10^2$. 
The charge moves from left to right along the line $(\pi-0)$.
}
\label{FULLCHE}
\end{figure}

We have considered the radiation 
produced by an electric charge 
moving at a constant velocity ${v \mathbf {\hat u}}$ 
perpendicular to the interface $z=0$ 
between two semi-infinite planar magnetoelectric media 
with the same refraction index.  
When $v$ is higher than $1/n$ 
we discover the emission of RVCR 
codified in the angular distribution illustrated in Fig.~\ref{FULLCHE}. 
The main characteristics of this RVCR are: 
(i) The threshold condition $v>1/n$ must be satisfied as in the standard case. 
(ii) The RVCR occurs for all frequencies in the \v{C}erenkov spectrum 
and it is always accompanied by the forward Vavilov--\v{C}erenkov radiation 
with the same frequency. 
(iii) The SD of the RVCR is  suppressed  with respect to the forward output 
by the factor ${\tilde \theta}^2/ (8 n^2)$.\cite{URRU} 

\section*{Acknowledgments}
OJF has been supported by the doctoral fellowship CONACYT-271523. 
OJF, LFU, and ORT acknowledge support from the projects: 
\#237503 from CONACYT and 
\# IN103319 from Direcci\'on General de Asuntos del Personal Acad\'emico (Universidad Nacional Aut\'onoma de M\'exico). 
We thank VAK and collaborators for a wonderful meeting.

\end{document}